\documentclass[12pt]{article}
\usepackage{epsfig,amsfonts,amssymb}
\usepackage{hyperref}
\pdfoutput=1

\usepackage{cite}
\usepackage{cite}

\usepackage{comment}

\def\ZZZ{{\hbox{ Z\kern-1.6mm Z}}}
\def\RRR{{\hbox{ R\kern-2.4mm R}}}
\def\CCC{{\hbox{ C\kern-2.0mm C}}}
\def\zzz{{\hbox{z\kern-1mm z}}}

\newcommand{\qeq}{{\hbox{=\kern-2.3mm ? \kern.5mm }}}
\renewcommand{\qeq}{=}

\newcommand{\be}{\begin{equation}}
\newcommand{\ee}{\end{equation}}
\newcommand{\ben}{\begin{eqnarray}\displaystyle}
\newcommand{\een}{\end{eqnarray}}

\newcommand{\refb}[1]{(\ref{#1})}
\newcommand{\p}{\partial}

\def\one{{\hbox{ 1\kern-.8mm l}}}
\def\zero{{\hbox{ 0\kern-1.5mm 0}}}

\newcommand{\bea}[1]{\begin{eqnarray}\label{#1} }
\newcommand{\eea}{\end{eqnarray}}

\newcommand{\eqref}{\refb}

\usepackage{bm}
\usepackage[table]{xcolor}

\begin{document}

\baselineskip 24pt

\begin{center}

{\Large \bf Are Moduli Vacuum Expectation Values or Parameters?}

\end{center}

\vskip .6cm
\medskip

\vspace*{4.0ex}

\baselineskip=18pt

\centerline{\large \rm Ashoke Sen}

\vspace*{4.0ex}

\centerline{\large \it International Centre for Theoretical Sciences - TIFR 
}
\centerline{\large \it  Bengaluru - 560089, India}


\vspace*{1.0ex}
\centerline{\small E-mail:  ashoke.sen@icts.res.in}

\vspace*{5.0ex}

\centerline{\bf Abstract} \bigskip

Banks has argued that the moduli of string theory are not vacuum expectation values
but parameters. We offer a different perspective on this question. Given two different
points P and Q in the moduli space, 
we shall regard them as different vacua of the same underlying
theory if in a theory where the asymptotic values of the moduli correspond to the
point P, we can perform an experiment that 
can determine the spectrum and S-matrix of a theory where the
asymptotic values of the moduli correspond to the point Q. We argue that in 
asymptotically flat space-time, this
can be achieved by taking a charged black hole in the limit of large mass and charges.
In this limit the local geometry at any point outside the horizon is indistinguishable from
flat space-time. However the moduli vary slowly over the entire region so that their
values at faraway  points can differ by order unity. Therefore, by sending out
experimental teams to different regions outside the horizon, an asymptotic observer
can measure the spectrum and S-matrix for different values of the moduli.

\vfill \eject

Many string vacua, particularly those with eight or more unbroken
supersymmetries, have moduli fields,
-- scalars with flat potential that can take any value at infinity\cite{2303.04819}. 
Thus different values of the
scalar fields label different possible string theories. The question that we shall address
is: do different asymptotic values of the scalar fields describe different string theories,
or are they to be regarded as different vacua of the same underlying string theory?
Banks has recently suggested that they should be regarded as different string 
theories\cite{2501.17697}.
We shall take a different perspective and argue that they should be regarded as
different vacua of the same underlying theory.

To address this question we need to first fix a criteria for distinguishing between these
two alternatives. We fix the following criterion. Suppose that P and Q are two different
points in the moduli space. We consider two theories, one 
where the asymptotic moduli take the value
corresponding to the point P and the other where the asymptotic 
moduli take the value
corresponding to the point Q. 
Let A be an observer in the first theory and B be an observer in the second theory.
Now the question we shall ask is: can the observer A  create a region in space
where the standard particle physics  experiments 
will yield the spectrum and S-matrix seen by the
observer B? If the answer is in the affirmative then we should consider the two
theories to be different vacua of the same theory since we are able to create the 
environment seen by one observer in the universe of the other observer. We shall
argue that this is the case for a wide family of points P and Q in the moduli space.

Our main tool in our analysis will be charged black holes.\footnote{One might worry
that superselection rules may prevent us from creating a charged black hole, but
this can be avoided by creating another black hole with opposite charge and
placing it far away.} A two derivative
theory with metric $g_{\mu\nu}$, 2-form field $B_{\mu\nu}$, $U(1)$ gauge fields
$\{A_\mu^{(i)}\}$ and moduli fields $\{\phi_\alpha\}$ in $D$ space-time
dimensions has a scaling `symmetry'
\be \label{escale}
g_{\mu\nu}\to \lambda^2\, g_{\mu\nu}, \quad B_{\mu\nu}\to \lambda^2\, B_{\mu\nu},
\quad A_\mu^{(i)}\to \lambda A_\mu^{(i)}, \quad \phi_\alpha\to \phi_\alpha\, ,
\ee
under which the action scales as
\be
S\to \lambda^{D-2} D\, .
\ee
This maps classical solutions to (inequivalent) classical solutions.
Under this transformation, 
the mass $M$, electric
charge $Q^{(i)}$ and angular momentum $J_m$ of a black hole solution 
scale as
\be  \label{escale2}
M\to \lambda^{D-3} M, \quad Q^{(i)}\to \lambda^{D-3} Q^{(i)}, \qquad J_m\to \lambda^{D-2}
J_m\, ,
\ee
and the Bekenstein-Hawking entropy scales as 
\be
S_{BH} \to \lambda^{D-2} S_{BH}\, .
\ee
The subscript $m$ of $J_m$ signifies that in general space-time 
dimensions there
may be more than one Cartan generators of the rotation group, giving different
angular momenta.

From the scaling laws \refb{escale} we can find the scaling of various invariant
scalars:
\be
R \sim \lambda^{-2}, \qquad g^{\mu\rho} g^{\nu\sigma} F^{(i)}_{\mu\nu}
F^{(j)}_{\rho\sigma} \sim \lambda^{-2}, \qquad g^{\mu\nu}
\p_\mu\phi_\alpha \p_\nu\phi_\beta \sim \lambda^{-2}, \qquad \hbox{etc}\, .
\ee
This shows that all the invariants approach zero for large $\lambda$ and hence
at generic point in the black hole background the space-time looks locally flat.

However the moduli fields do not take the same value everywhere in the
black hole background. To see this note that in this limit the size of the black hole
scales as $\lambda$. Therefore, even though the norm of $\p_\mu\phi$ is of
order $\lambda^{-1}$, over a distance scale of order $\lambda$, $\phi_\alpha$ changes
by order unity. Therefore observers placed at different points in the black hole
background will see different values of the moduli and  by performing local 
experiments, these different observers can determine the spectrum and S-matrix of the
theory for different values of the moduli. As long as the size of the experimental
set-up and the energies remain 
finite (or increase at a rate slower than $\lambda$) 
as we take the large $\lambda$ limit,  no effect of background
gauge fields, curvature or scalar field gradient is felt in these experiments.
The result of these experiments can then be 
transmitted to the asymptotic observer, who will therefore have access to the 
results for a wide set of points in the moduli space. Theories corresponding to these
different values of the moduli should then be regarded as different vacua of a single
theory.

While this argument shows that some points in the moduli space can be related
to each other
this way, it does not show that all points can be related. Indeed there are
examples where this is not true, {\it e.g.} in ten dimensional type IIB string theory
we have axion-dilaton moduli but no charged black holes since there are no gauge
fields. However, black holes are not the only objects that can be used for this
purpose.  We can use a black string (or black $p$-branes). 
The scaling laws \refb{escale} and \refb{escale2} extend easily to the case of
$k$-form gauge fields $C^{(k)}$ and $p$-brane charges $Q^{(p)}$ as follows:
\be \label{escale3}
C^{(k)}_{\mu_1\cdots \mu_k} \sim \lambda^k \, C^{(k)}_{\mu_1\cdots \mu_k},
\qquad Q^{(p)}\to \lambda^{D-3-p} \, Q^{(p)}\, .
\ee
A large loop of  black string or a large black brane of spherical topology  will
also have space-time varying moduli and in the scaling limit \refb{escale},
\refb{escale3} the
space-time around any point will look locally flat except that the moduli will take 
different values at different points in space-time. In particular even though such 
configurations will be time dependent, in the large $\lambda$ limit
the time dependence will be invisible to an
experiment that is performed within a finite time. We can also take multiple branes
of various kinds, including merging black holes whose merger time will scale as
$\lambda$. While we have not proved that by
taking all such brane configurations we can explore all points in the moduli space,
it is clear that we can access a wide class of points in the moduli space in this way.

We shall end this note by giving a simple example where we can see the varying
moduli. We take a black hole solution carrying D0-brane charge in ten dimensional
type IIA string theory. The solution was written down in \cite{horowitz} and the string
metric $ds^2$ and the dilaton $\phi$ take the
form:
\ben
ds^2 &=& - \left[ 1 - \left({r_+\over r}\right)^7\right] 
\left[ 1 - \left({r_-\over r}\right)^7\right]^{-1/2} dt^2 
+ \left[ 1 - \left({r_+\over r}\right)^7\right] 
\left[ 1 - \left({r_-\over r}\right)^7\right]^{-17/14} dr^2 \nonumber \\
&& + r^2 \, \left[ 1 - \left({r_-\over r}\right)^7\right]^{-3/14} d\Omega_8^2, \nonumber \\
e^{-2\phi} &=& \left[ 1 - \left({r_-\over r}\right)^7\right]^{3/2}\, ,
\een
where $d\Omega_8^2$ is the metric on the unit eight sphere and $r_\pm$ with $r_-<r_+$
are the parameters that control the mass and charge of the black hole. The outer horizon
is situated at $r=r_+$. The scaling described in \refb{escale}, \refb{escale2}
corresponds to scaling $r_\pm$, $r$ and $t$ by $\lambda$. In this limit all invariant scalars
constructed out of two derivatives of the fields scale as $\lambda^{-2}$. Therefore the
space-time is locally flat for large $\lambda$. The dilaton 
remains invariant under this scaling and span the range
\be
\left[ 1 - \left({r_-\over r_+}\right)^7\right]^{3/2} < e^{-2\phi} < 1\, ,
\ee
as $r$ varies from $r_+$ to $\infty$. In particular by taking the extremal limit 
$r_-\to r_+$, we can extend the range to $0 < e^{-2\phi} < 1$. By doing experiments
at different radial distances from the horizon, the asymptotic observer could measure
the spectrum and the S-matrix for different values of the string coupling.

Note that this configuration allows us to find regions where $\phi$ is positive,
To find negative $\phi$ regions in space-time, one could consider the
vicinity of large spherical black four branes or six branes.

\bigskip

\noindent{\bf Acknowledgement:} 
I thank Tom Banks for useful discussions.
This work was supported by the ICTS-Infosys Madhava 
Chair Professorship, the J. C. Bose fellowship of the Department of Science
and Technology, India
and the Department of Atomic Energy, Government of India, under project no. RTI4001.

\end{document}